\documentclass[final,5p,times,twocolumn]{elsarticle}

\usepackage{lineno}
\usepackage{amssymb}
\usepackage{graphicx}
\usepackage{booktabs}
\usepackage{textcomp}
\usepackage[thinspace,noams,squaren]{SIunits}
\usepackage{url}
\usepackage{setspace}

\begin{document}
\begin{frontmatter}

\title{Low-energy \textmu\ensuremath{^{+}}\ via frictional cooling}

\author[cas,mpp]{Yu Bao\corref{cor1}}
\ead{baoyu@ihep.ac.cn}
\author[mpp]{Allen Caldwell}
\author[mpp]{Daniel Greenwald}
\author[mpp]{Guoxing Xia}

\address[cas]{Institute of High Energy Physics, Chinese Academy of Sciences, Beijing, China}
\address[mpp]{Max Planck Institute for Physics, Munich, Germany}

\cortext[cor1]{Corresponding author}

\begin{abstract}
	Low-energy muon beams are useful for a range of physics experiments. We consider the production of low-energy \textmu\ensuremath{^{+}}\ beams with small energy spreads using frictional cooling. As the input beam, we take a surface muon source such as that at the Paul Scherrer Institute. Simulations show that the efficiency of low energy \textmu\ensuremath{^{+}}\ production can potentially be raised to 1\%, which is significantly higher than that of current schemes.
\vspace{1pc}
\end{abstract}


\end{frontmatter}
\section{Introduction}
	Intense low-energy muon (LE-\textmu\ensuremath{^{+}}) sources with small energy spreads are important for many experiments including measurements of rare muon decays, fundamental constants, and the muon's anomalous magnetic moment; determination of the Lorentz structure of the weak interaction; and tests of CPT symmetry and QED. They are also useful in condensed matter physics: muon beams with small  energy spreads and mean energies tunable from  eV to keV can be used as sensitive probes to study various phenomena in materials~\cite{low_mu_exp}. A muon collider~\cite{allen2} and neutrino factory~\cite{neutrino_fac} also require high-quality muon beams.
	
	Muon beams are tertiary beams, produced in the decay of pion beams that are produced by proton beams hitting targets. They initially have large energy spreads that must be reduced. This process, called ``cooling,'' is important for all the experimental endeavors described above. Frictional cooling is one promising method to reduce the energy spread. 
	
	In this paper, we explain the frictional cooling concept and detail a scheme for the production of a LE-\textmu\ensuremath{^{+}} beam from a surface muon source, such as the one available at the Paul Scherrer Institute (PSI), Switzerland~\cite{PSI_par}. We then compare the efficiency of our cooling scheme to the current LE-\textmu\ensuremath{^{+}} production setup at PSI~\cite{PSI_sMu}.

\section{Frictional Cooling}
\begin{figure}[!t]
\centering
	\includegraphics[width=\columnwidth]{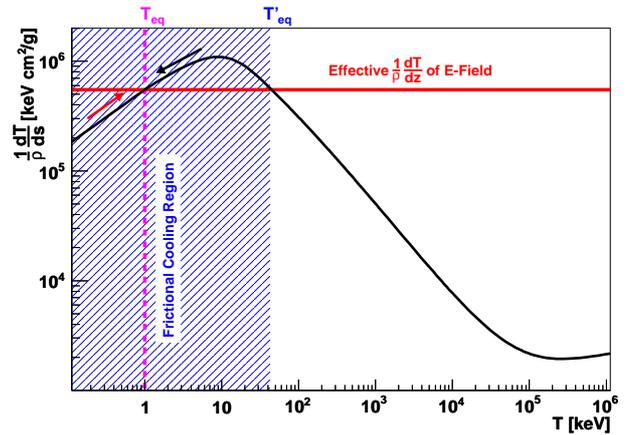}
	\caption{Stopping power for \textmu\ensuremath{^{+}}\ in helium as a function of kinetic energy $T$. $T_\mathrm{eq}$ is the equilibrium energy. The red line represents the accelerating power of an external electric field. The shadowed band is the frictional cooling region.\label{dEdx}} 
\end{figure}
Frictional cooling is the bringing of charged particles to an equilibrium energy by the balancing of energy loss to a material with energy gain from an electric field~\cite{Muhlbauer}. Figure~\ref{dEdx} shows the stopping power, $\frac{1}{\rho}\,\frac{\mathrm{d} T}{\mathrm{d} s}$, as a function of kinetic energy $T$ for \textmu\ensuremath{^{+}}\ , where $\rho$ is the density of the medium and $\frac{\mathrm{d} T}{\mathrm{d} s}$ is the energy loss per unit path length. The stopping power has been calculated using velocity-scaled proton data~\cite{NIST}. Applying an electric field to restore kinetic energy in the longitudinal direction brings the particles to the equilibrium energy ($T_\mathrm{eq}$). Particles with kinetic energies less than $T_\mathrm{eq}$ accelerate because they gain more energy from the electric field than they lose to the material. Particles with kinetic energies greater than $T_\mathrm{eq}$ but less than $T_\mathrm{eq}'$ (see Fig.\ \ref{dEdx}) decelerate because they lose more energy than they gain. Thus all particles with kinetic energy less than $T_\mathrm{eq}'$ approach $T_\mathrm{eq}$ and the energy spread decreases. 

Frictional cooling requires particles have low kinetic energies. Near $T_{eq}$, where particles lose energy  through excitations of the material atoms, charge exchange processes, and nuclear recoil, the stopping power of the retarding medium is large. The average density must be kept low so that $\frac{\mathrm{d} T}{\mathrm{d} s}$ can be balanced by realistic electric fields.

One way to achieve this is using a series of thin foils with an electric field present between them, as was used by a previous experiment at PSI~\cite{Muhlbauer}. Although the average density is low, the density in the foils is large, so the lower-energy particles have a significant probability to stop and decay. At low kinetic energies, a positively-charged particle picks up and loses electrons as it passes through the material~\cite{senba}. With each passage of the beam through a foil, a fraction of the beam exits in the neutral charge state (muonium). The muonium atoms are not reaccelerated by the electric field and have small probabilities to survive. In this way, a large portion of the beam neutralizes and the efficiency is low.

We avoid both problems by using gas as the low-density retarding medium. Energy is lost and restored over very small distances, so that the energy of the \textmu\ensuremath{^{+}} fluctuates about $T_\mathrm{eq}$. Charge exchange processes still proceed in the gas, but muonium atoms change back to \textmu\ensuremath{^{+}} over short distances, so the \textmu\ensuremath{^{+}} are not lost but have a reduced effective charge. This decreases the energy gain from the electric field. To minimize this decrease, we use helium gas because in it the effective charge is largest~\cite{nakai}. 

Elastic scattering off of nuclei places a lower limit on the energy spread achievable by frictional cooling~\cite{allen2}. The energy spread of the cooled beam can be preserved during reacceleration, yielding higher energy beams with small relative energy spreads.
 
\section{Cooling scheme}
	We consider producing a LE-\textmu\ensuremath{^{+}} beam using frictional cooling from a surface muon source such as the \textmu E4 beam at PSI.

\subsection{LE-\textmu\ensuremath{^{+}} beam at PSI}
	The \textmu E4 beam line delivers the world's highest flux surface \textmu\ensuremath{^{+}} beam~\cite{PSI_par}. PSI currently produces a LE-\textmu\ensuremath{^{+}} beam by focusing the \textmu E4 beam onto a thin foil moderator (about \unit{100}{\micro\meter} thick) covered with a very thin layer (less than \unit{1}{\micro\meter}) of a condensed van der Waals gas such as argon, neon, or nitrogen cryosolids. Very slow \textmu\ensuremath{^{+}}\ exit from the downstream side of the moderator. The energy distribution of these muons has a maximum near \unit{15}{\electronvolt} and a tail extending to higher energies. An electric field then reaccelerates the muons yielding a beam with mean energy tunable between \unit{0.5}{\kilo\electronvolt} to \unit{30}{\kilo\electronvolt} with a spread of \unit{400}{\electronvolt}. This moderation technique has an efficiency of $10^{-5}$ to $10^{-4}$ to convert a surface muon to a LE-\textmu\ensuremath{^{+}}~\cite{PSI_sMu}. 

\subsection{Frictional cooling scheme}
\begin{figure}[!h]
   \centering
	\includegraphics[width=\columnwidth]{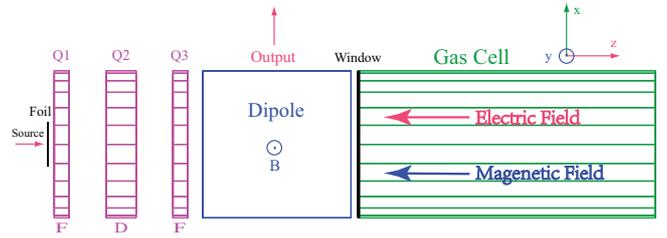}
	\caption{The block diagram of the frictional cooling scheme. \label{scheme_PSI}}
\end{figure}

	We searched for a way to efficiently produce a LE-\textmu\ensuremath{^{+}} beam using frictional cooling. Because frictional cooling only works at low energies, we must lower the kinetic energies of the muons to below \unit{50}{\kilo\electronvolt}. To achieve this, we inject the muons into a cooling cell, which is a helium gas cell with an electric field in the direction opposite to that of the beam (Fig.\ \ref{scheme_PSI}). A \unit{1}{\tesla} solenoidal magnetic field guides the beam in the cooling cell. The gas and the electric field slow the muons down. The slowing down is primarily due to the electric field. When they turn back, a large fraction of the muons have very low kinetic energies and are brought to $T_\mathrm{eq}$. This scheme produces a beam at the equilibrium energy with a very small  energy spread, regardless of the initial beam energy spread. 	

	To reduce the required electric potential in the cooling cell, the beam first passes through a foil for an initial energy loss. For a fixed electric field, lowering the mean energy decreases the distance that the muons traveled in the cooling cell, thereby reducing the loss from decay.
	
	The cooling cell is ineffective for muons with large transverse momentum, because when they turn around they have kinetic energies above $T_\mathrm{eq}'$ (see Fig.\ \ref{dEdx}). The electric field reaccelerates them to approximately their initial energies. Therefore we use quadrupoles to reduce the angular divergence of the beam. The quadrupoles are designed to maximize the number of muons with low transverse momenta.
	
	Between the quadrupoles and the cooling cell is a weak dipole. It has minimal effect on the high-energy input beam, letting it pass straight through. After the dipole, the beam enters the cooling cell. 
	
	A window is needed to separate the cooling cell from the input beam line. The window must be thin to allow the cooled muons get out efficiently. After the muons exit the cooling cell, they enter the dipole again. The magnetic field in the $y$ direction turns the cooled muons in the $x$ direction. We evaluate the momenta and positions of the muons as they exit the dipole.

\section{Simulation}
 \subsection{Simulation tools}
	We have developed a program called CoolSim~\cite{coolsim} based on Geant4~\cite{g4}, in which geometries can be implemented easily by a macro-command interface. For simulating the frictional cooling process, we use the Geant4 low-energy processes~\cite{g4low} to determine energy loss and spatial displacement. 
	
\begin{figure}[ht]
	\centering
	\includegraphics[width=\columnwidth]{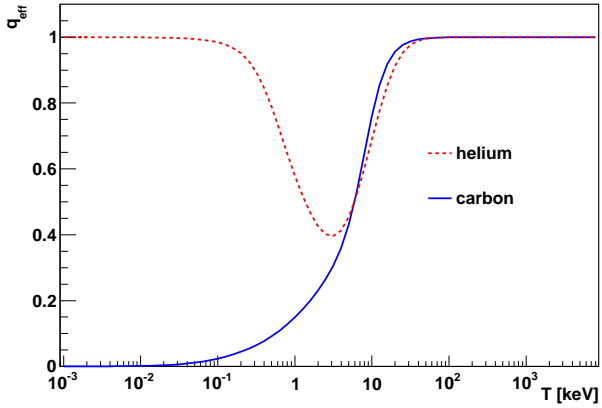}
	\caption{The effective charge as a function of kinetic energy for \textmu\ensuremath{^{+}}\ in helium (dashed) and carbon (solid). \label{eff_charge}}
\end{figure}

	We have added new low-energy physics processes to the Geant4 framework: hydrogen formation for protons, muonium formation for \textmu\ensuremath{^{+}}, and charge exchange interactions for both protons and \textmu\ensuremath{^{+}}. The charge exchange process is simulated by assigning particles an effective charge,
	\begin{displaymath}
		q_{\mathrm{eff}} = \frac{\sum q_{j}\sigma_{ij}}{\sum\sigma_{ij}}, \nonumber
	\end{displaymath}
	where $\sigma_{ij}$ is the cross section for the changing charge state from $q_{i}$ to $q_{j}$ \cite{deg}. The effective charges for \textmu\ensuremath{^{+}}\ in helium and carbon are shown in Fig.\ \ref{eff_charge}.
	
	We also take the decay process for \textmu\ensuremath{^{+}}\ into account. The energy losses calculated by the program were verified against the NIST tables.

	We simulated the scheme shown in Fig.\ \ref{scheme_PSI} in the CoolSim program using uniform electric and magnetic fields without fringe fields. The quadrupoles are simulated as a slightly modified FODO cell, which is a symmetric $\frac{1}{2}FODO\frac{1}{2}F$ cell. All components have the same aperture. The simulation parameters are given in Table\ \ref{scheme_par}.
\begin{table}[hbt]
   \centering
   \caption{Parameters of the elements in the cooling scheme}
   \begin{tabular}{ll}
       \toprule
       \midrule
       	Aperture 		&\unit{100}{\centi\meter} \\
	\hline
       	Foil:  \\
	Thickness			 & \unit{100}{\micro\meter}	\\
	Material			&tungsten	\\
	\hline
          Quadrupoles: \\
          Gradient         & \unit{1}{\tesla\per\meter}        \\
   	 Magnet Length             & \unit{20}{\centi\meter}             \\
	Drift space                             & \unit{40}{\centi\meter}          \\
	\hline
	Dipole: \\
	Strength                    &\unit{0.003}{\tesla}		\\
	Length 	&\unit{100}{\centi\meter} \\
	\hline
 	Cooling cell:	\\
	Length	&\unit{200}{\centi\meter}	\\
	Electric field strength	&\unit{1.8}{\mega\volt\per\meter}\\
	Magnetic field strength 	&\unit{1}{\tesla}\\
	Gas material			&helium	\\
	Gas density			&\unit{0.01}{\milli\gram\per\centi\cubic\meter}\\
	\hline
	Window:\\
	Thickness			&\unit{20}{\nano\meter}\\
	Material			&carbon \\
       \bottomrule
   \end{tabular}
   \label{scheme_par}
\end{table}

\subsection{Input beam}
\begin{table}[hbt]
   \centering
   \caption{Summary of \textmu E4 beam properties at PSI}
   \begin{tabular}{lcc}
       \toprule
       \midrule
       	Beam energy			 & \unit{3.7}{\mega\electronvolt}	\\
          Accepted solid angle         & \unit{135}{\milli\steradian}        \\
   	$x/x^{\prime}$ (FWHM)           & \unit{6.5}{\centi\meter}/\unit{150}{\milli\radian}   \\
	$y/y^{\prime}$ (FWHM)              & \unit{2.6}{\centi\meter}/\unit{300}{\milli\radian}   \\
	$\Delta$p/p (FWHM)	            & 9.5\%                              \\
	\textmu\ensuremath{^{+}} rate       & \unit{228\times\power{10}{6}}{\per\milli\ampere.\second}  \\
       \bottomrule
   \end{tabular}
   \label{Par_PSI}
\end{table}

\begin{figure}[ht]
	\centering
	\includegraphics[width=\columnwidth]{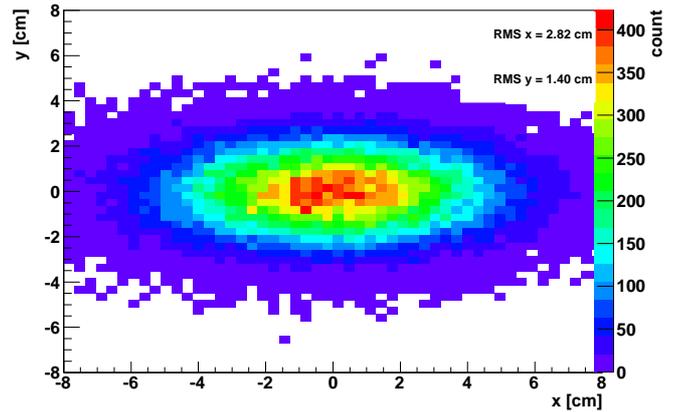}
	\caption{The spatial distribution of the simulated source. \label{Source}}
\end{figure}
\begin{figure}[ht]
	\centering
	\includegraphics[width=\columnwidth]{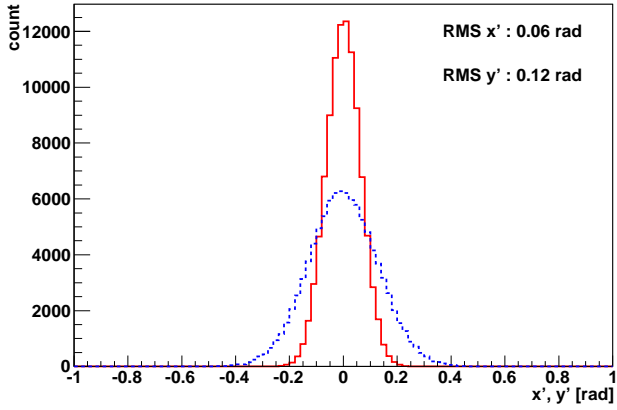}
	\caption{The angular distributions of the simulated source. \label{Angle}}
\end{figure}
We use the Geant4 General Particle Source to reproduce the surface muon beam using the parameters of PSI's \textmu E4 beam (Table\ \ref{Par_PSI}) and assuming gaussian distributions with no correlations. Figure\ \ref{Source} shows the spatial distribution of the simulated source, which agrees well with reference \cite{PSI_par}. The RMS of the simulated angular distributions are \unit{0.06}{\radian} in $x$ and \unit{0.12}{\radian} in $y$ (Fig.\ \ref{Angle}). The kinetic energy distribution of the simulated source has a mean value around \unit{3.7}{\mega\electronvolt} with an RMS of \unit{0.32}{\mega\electronvolt} (see Fig.\ \ref{Foil} solid curve). 100,000 muons were simulated.

\subsection{Effect of the foil}
\begin{figure}[ht]
	\centering
	\includegraphics[width=\columnwidth]{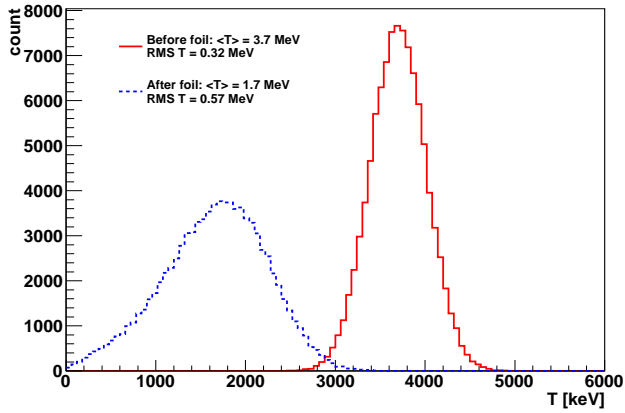}
	\caption{The kinetic energy distribution before (solid) and after (dashed) the foil. \label{Foil}}
\end{figure}
	The tungsten foil reduces the mean of the energy distribution of the beam from \unit{3.7}{\mega\electronvolt} to \unit{1.7}{\mega\electronvolt} (Fig.~\ref{Foil}) with 90\% efficiency. Thicker foils greatly reduce the efficiency, while only reducing the mean energy slightly. The angular spreads of the beam increase greatly after passing through the foil, which makes the quadrupoles more necessary.

\subsection{Effect of the quadrupoles}
\begin{figure}[ht]
	\centering
	\includegraphics[width=0.48\columnwidth]{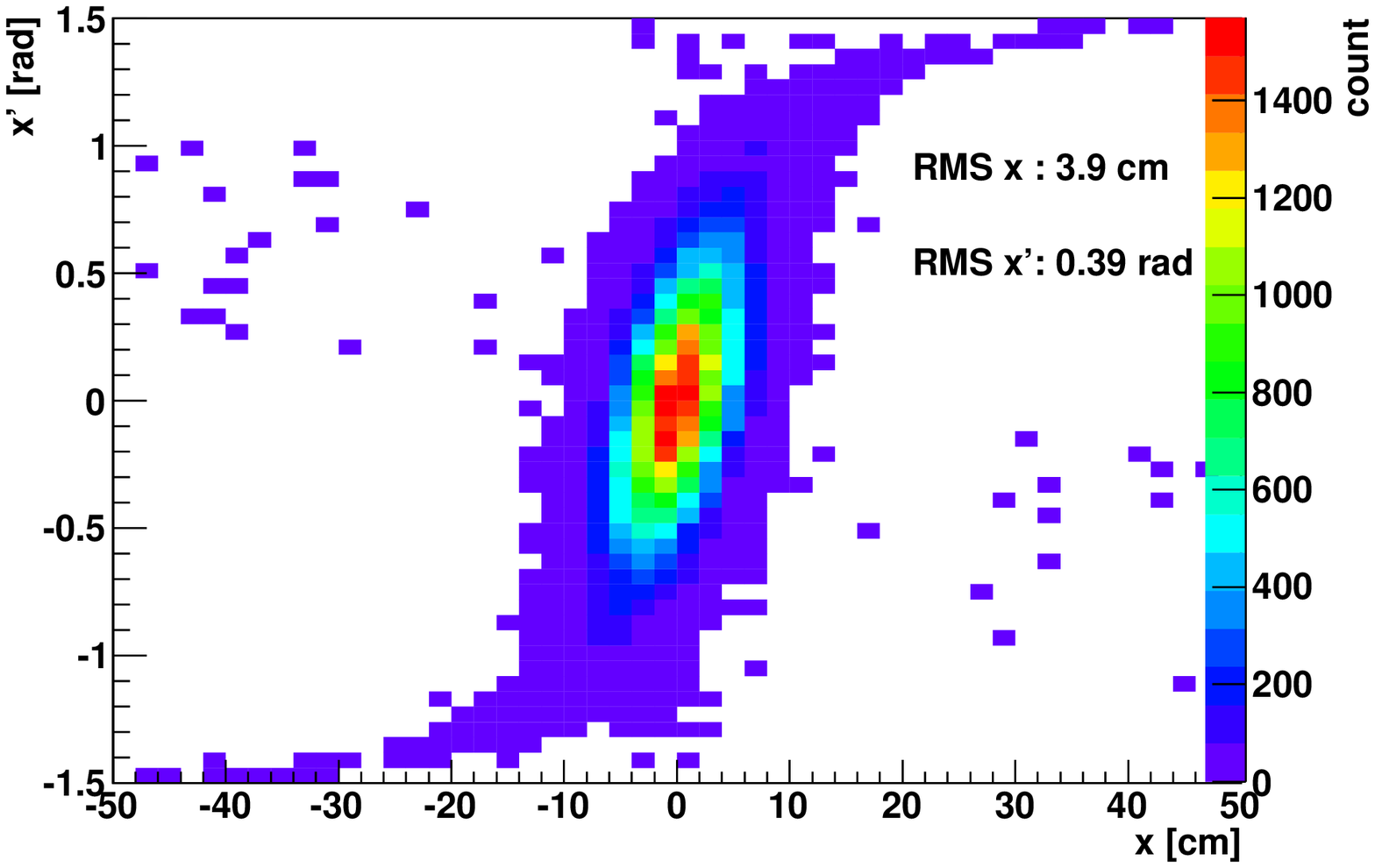}
	\includegraphics[width=0.48\columnwidth]{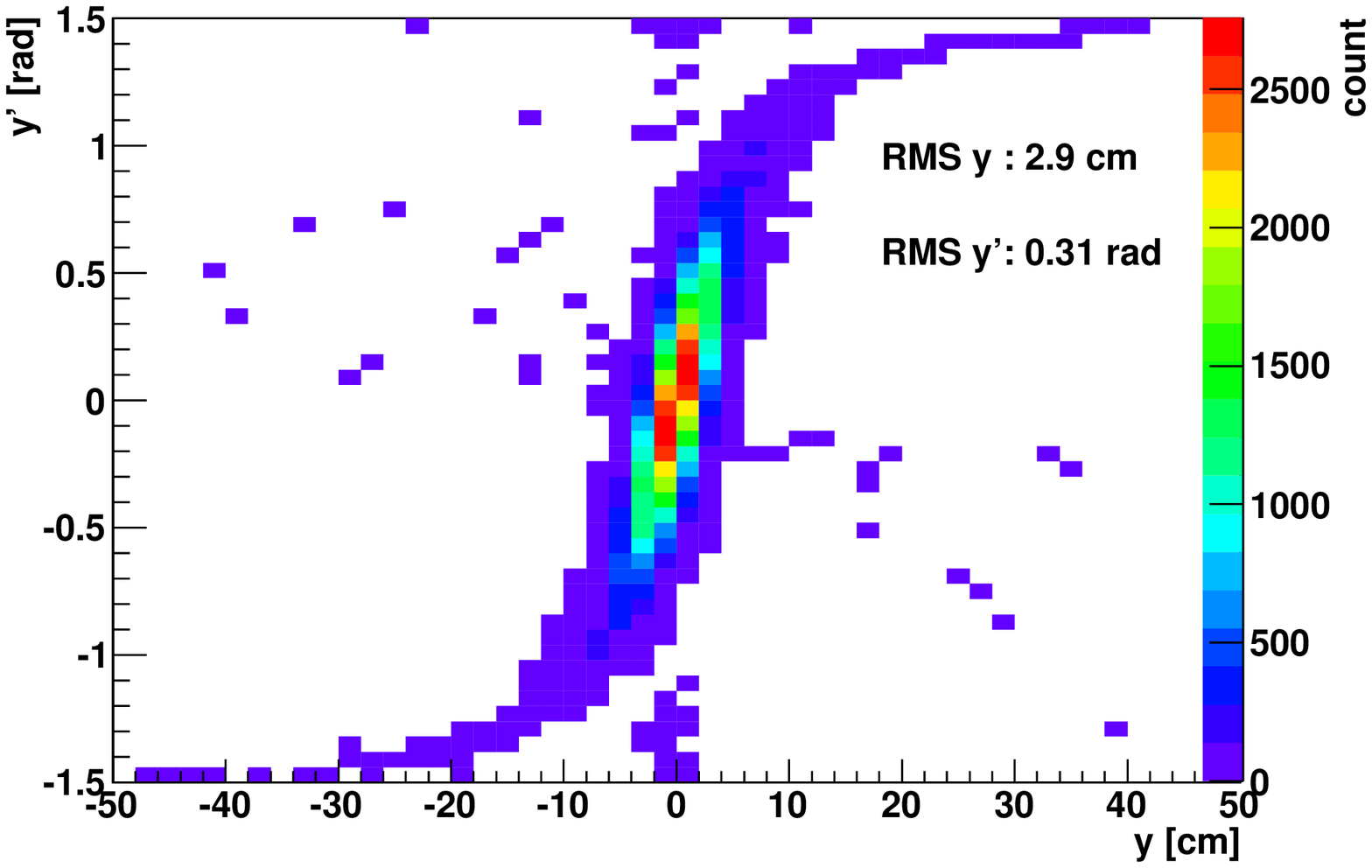}
	\includegraphics[width=0.48\columnwidth]{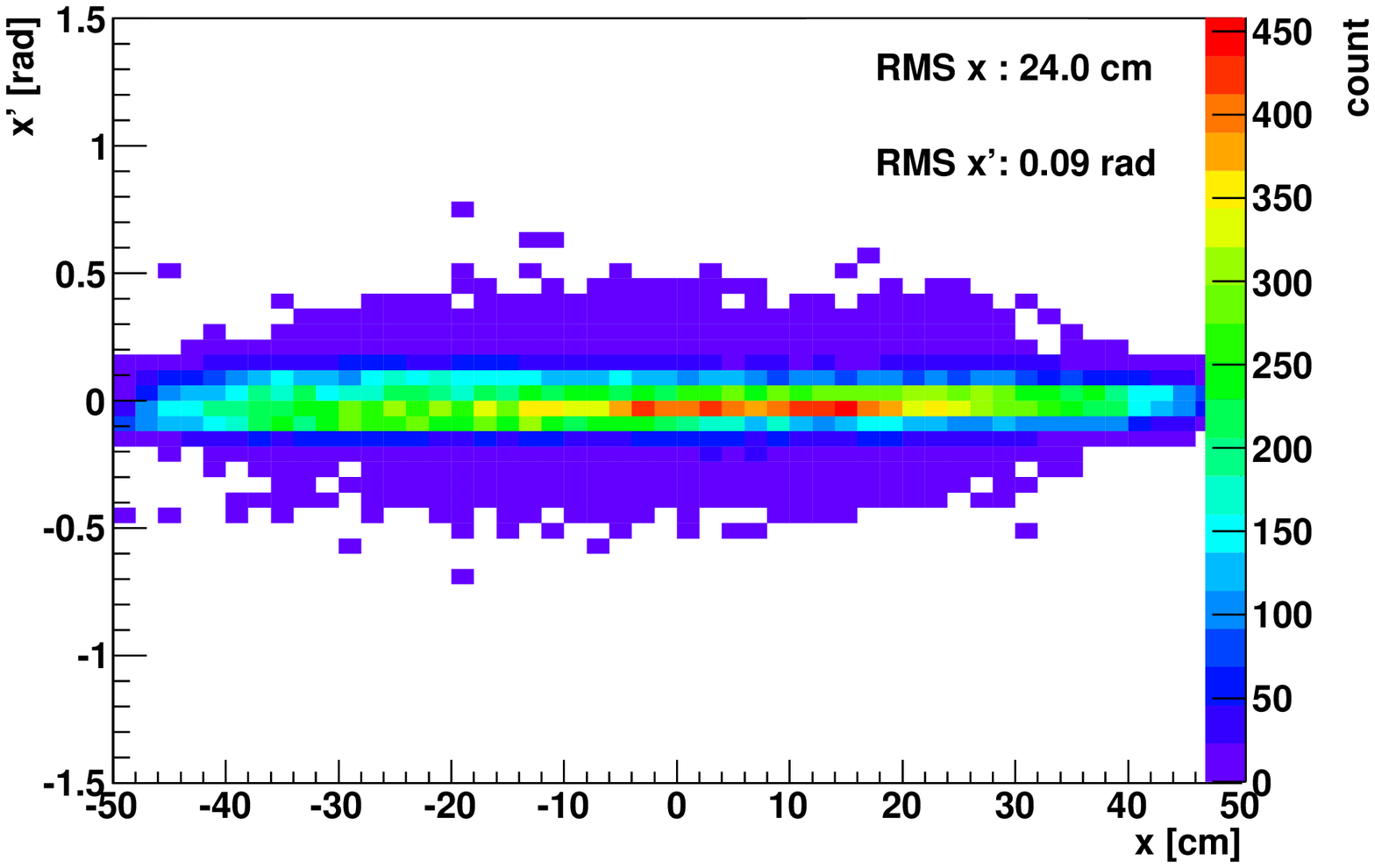}
	\includegraphics[width=0.48\columnwidth]{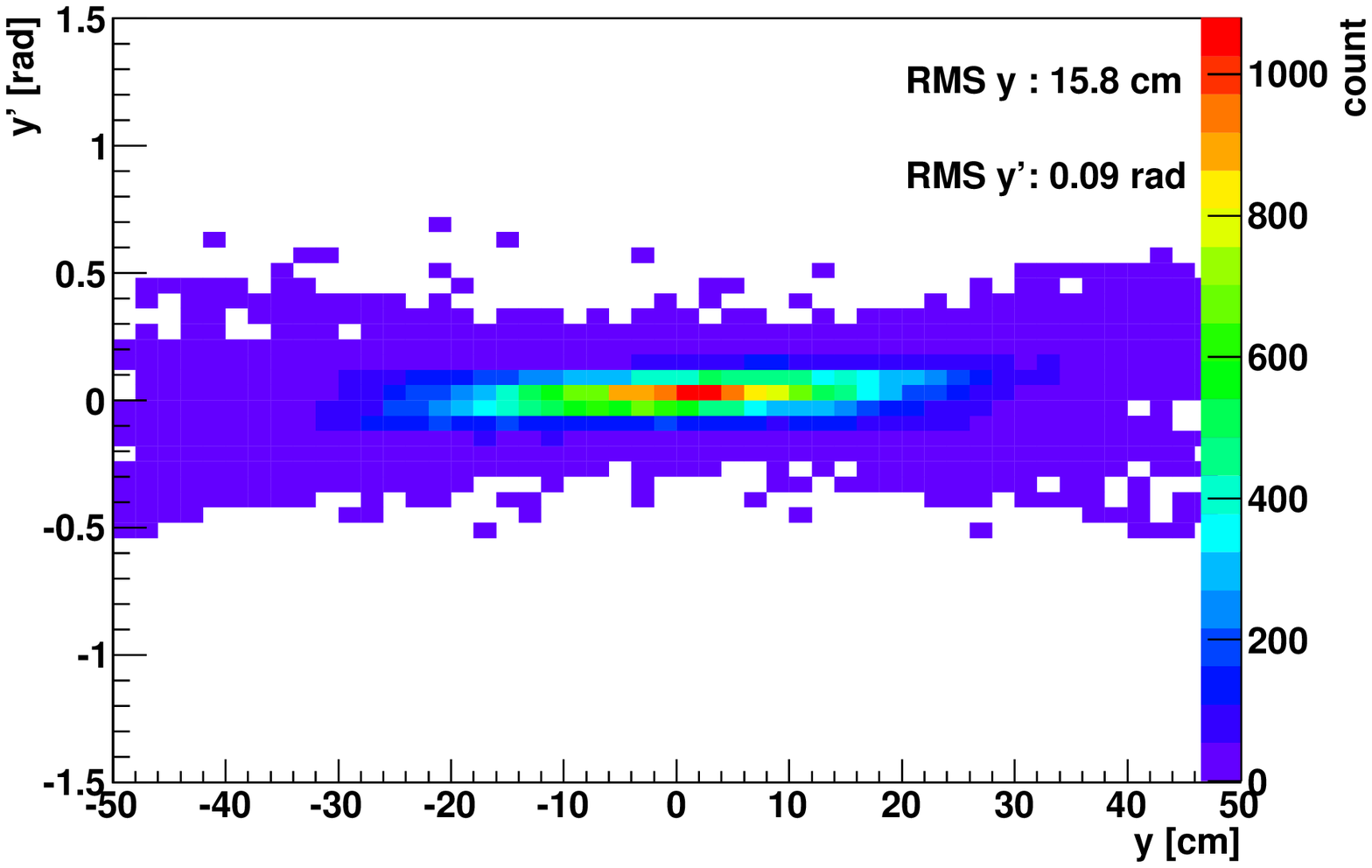}
	\caption{The phase spaces in $x/x^{\prime}$ (left)  and $y/y^{\prime}$ (right) directly before quadrupoles (upper) and directly before the cooling cell (lower).\label{phase}}
\end{figure}

The quadrupoles have been designed to rotate the beam phase space by \unit{90}{\degree} (Fig.\ \ref{phase}), enlarging the beam's transverse size and increasing the number of muons with low transverse kinetic energies. 

\begin{figure}[ht]
	\centering	
	\includegraphics[width=0.48\columnwidth]{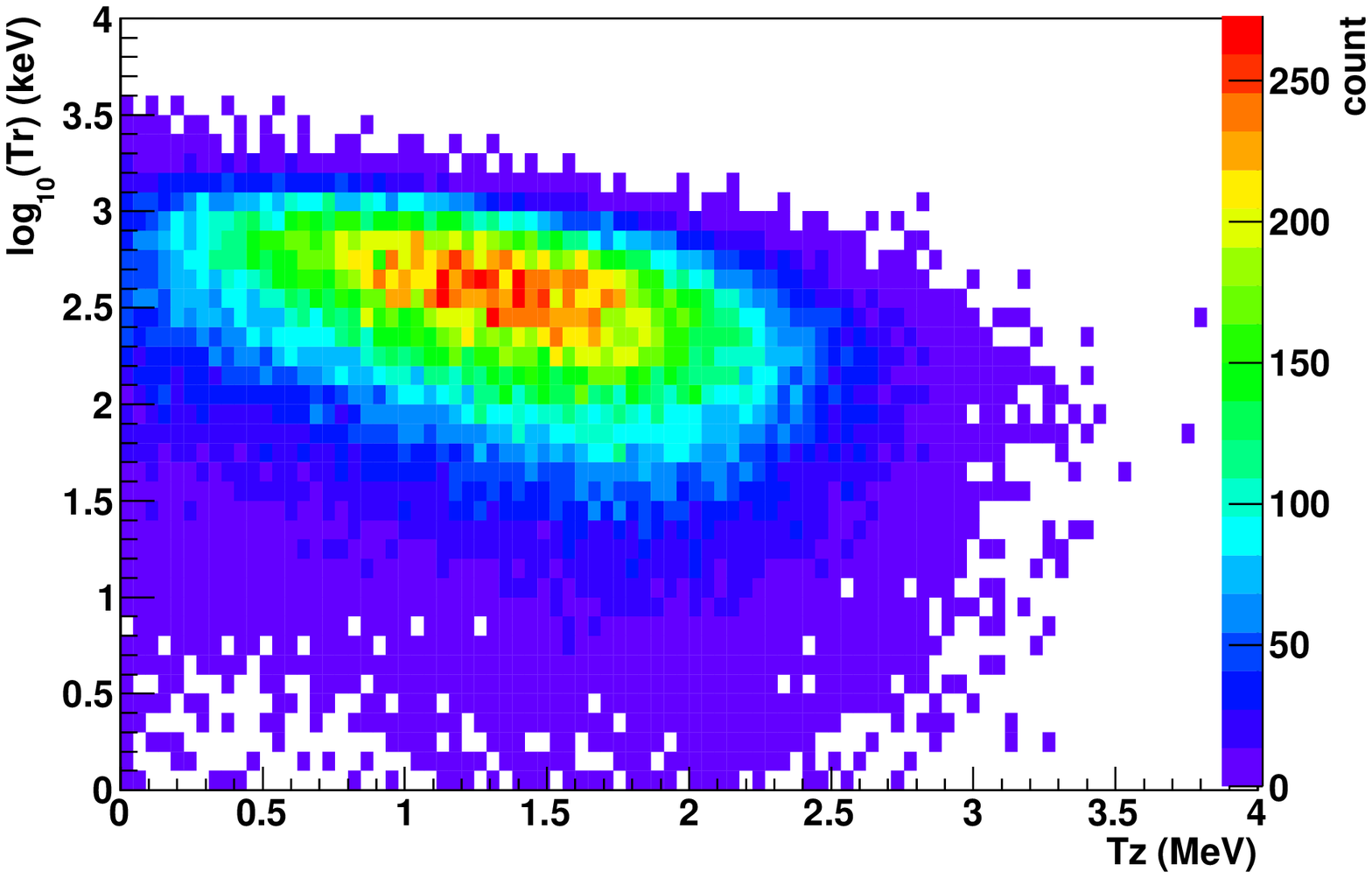}
	\includegraphics[width=0.48\columnwidth]{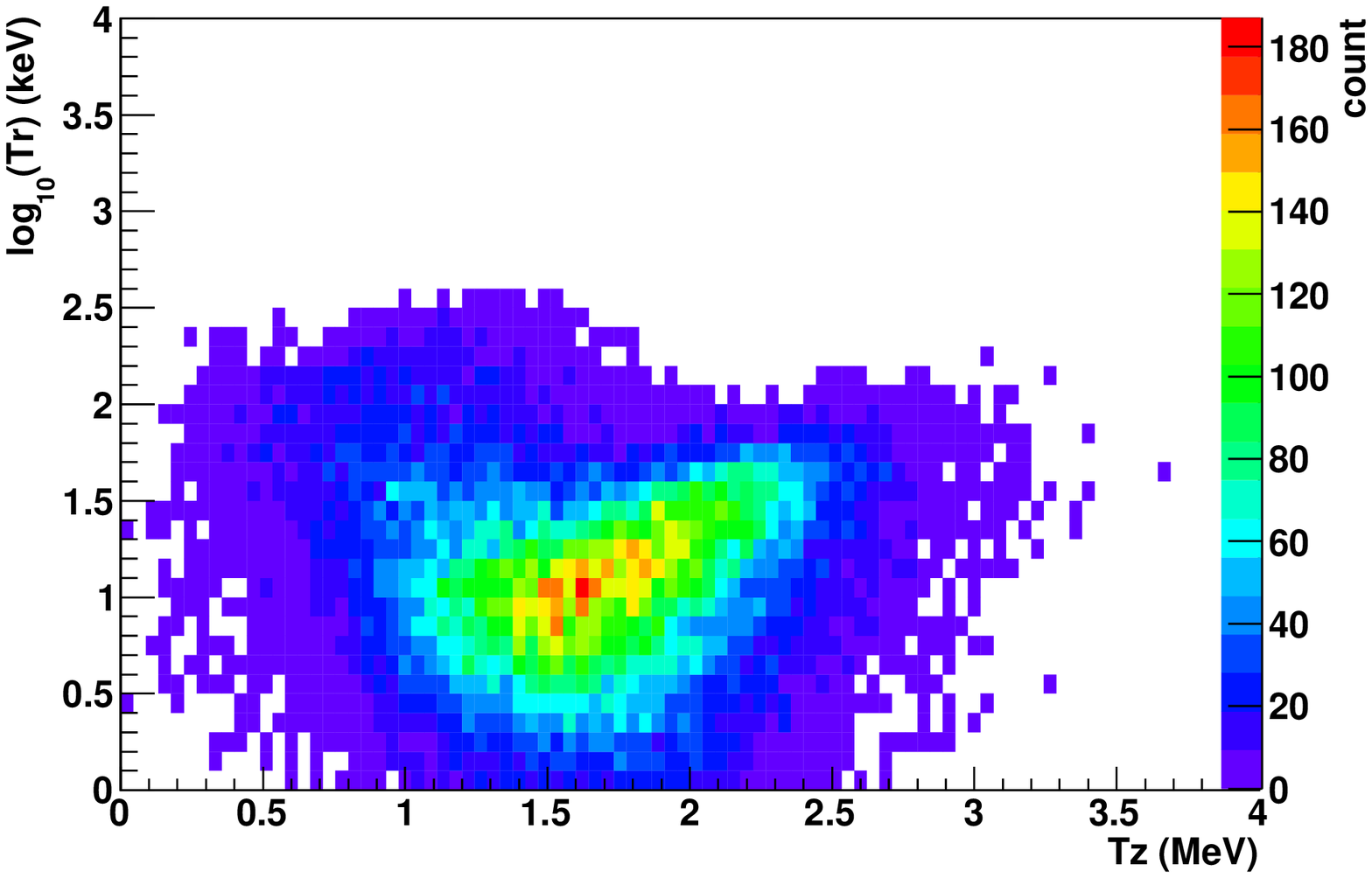}
	\includegraphics[width=0.48\columnwidth]{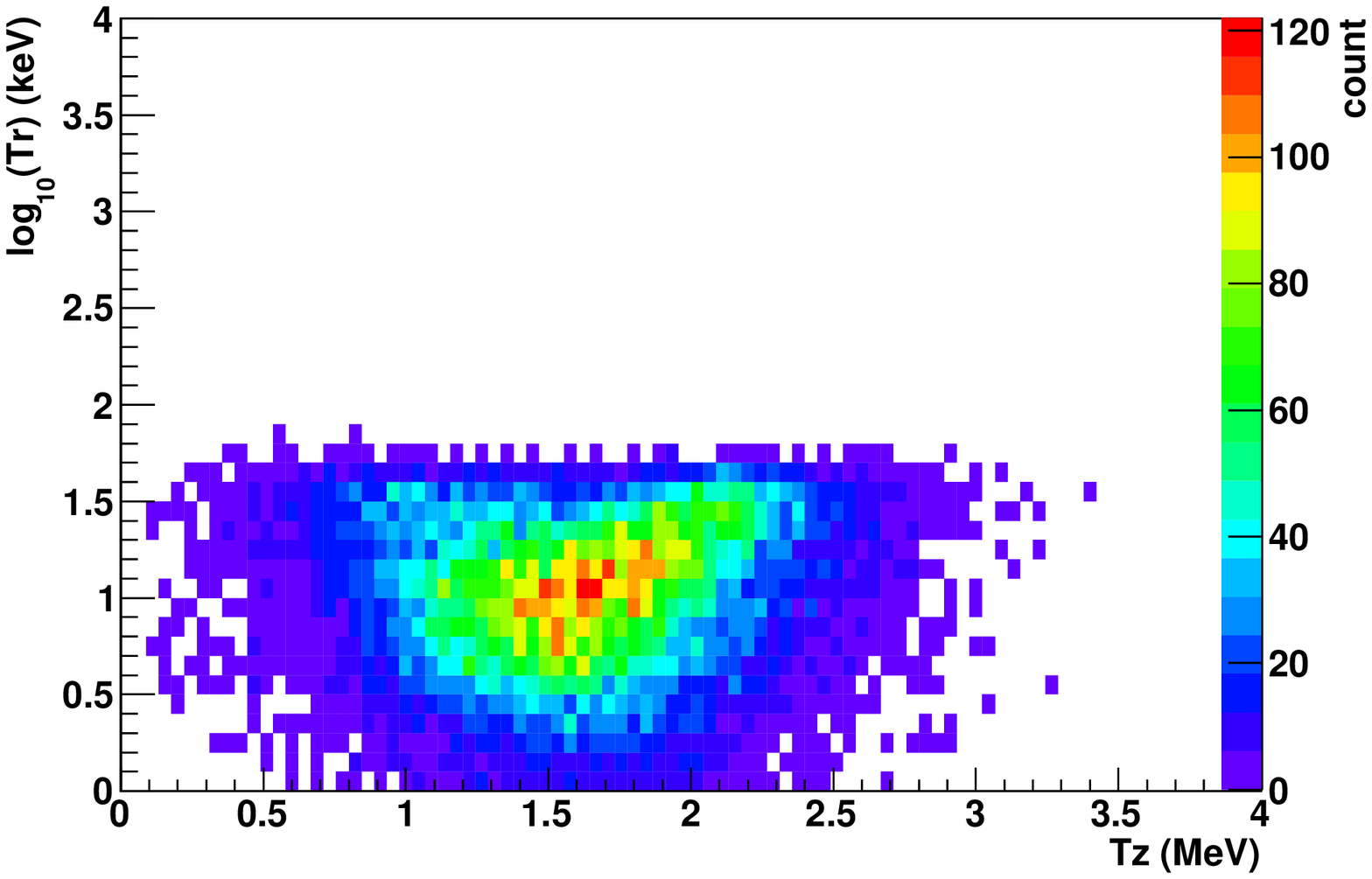}
	\includegraphics[width=0.48\columnwidth]{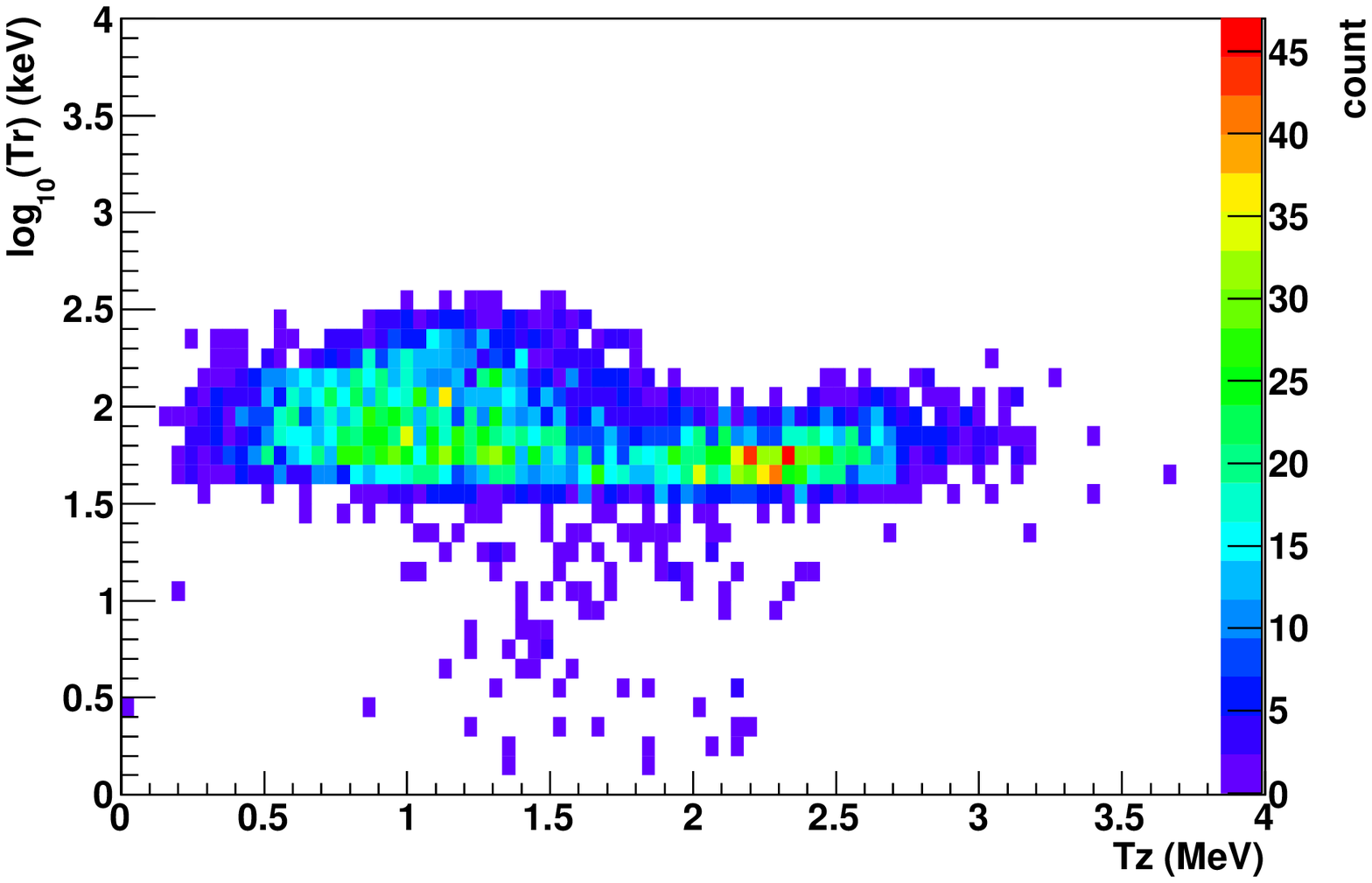}
	\caption{The transverse ($T_{r}$) and longitudinal ($T_{z}$) kinetic energy distributions of the beam directly before the quadrupoles (upper left) and directly before the cooling cell (upper right). The lower plots show the distributions directly before the cooling cell for the subset of muons that are ultimately cooled (left) and that are not cooled (right).\label{KET_Compare}}
\end{figure}

Figure\ \ref{KET_Compare} shows the longitudinal and transverse kinetic energy distributions of the beam directly before the quadrupoles and directly before entering the cooling cell. The mean transverse kinetic energy of the beam is greatly reduced. Figure\ \ref{KET_Compare} also shows the kinetic energy distribution before entering the cooling cell for the subset of the beam that can ultimately be cooled and for the subset that cannot be cooled. Muons with transverse kinetic energies greater than \unit{50}{\kilo\electronvolt} are not cooled. 

Half of the muons exiting the foil are lost due to the limit of the aperture of the quadrupoles. 

\subsection{Behavior in the cooling cell}	
 \begin{figure}[ht]
	\centering
	\includegraphics[width=\columnwidth]{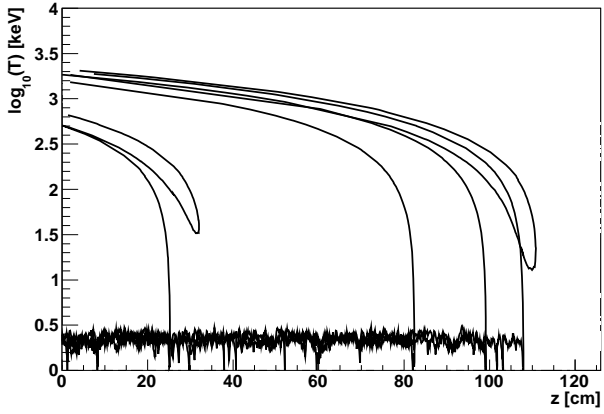}
	\caption{The kinetic energy as a function of depth in the cooling cell.  z = \unit{0}{\centi\meter} is the entrance of the cooling cell for representative trajectories. \label{KEZ}}
\end{figure}

	Figure\ \ref{KEZ} shows the kinetic energy of representative muons as a function of the depth in the cooling cell. The muons enter the cooling cell at z = \unit{0}{\centi\meter} with a mean kinetic energy of \unit{1.7}{\mega\electronvolt}. They are decelerated by the electric field and the gas and drift back out towards the entrance of the cell. Four of the muons shown in Fig.\ \ref{KEZ} reach $T_{eq}$, while two turn back and are reaccelerated. The fluctuations in the kinetic energies of the cooled muons about $T_\mathrm{eq}$ are due to large angle scatters off of gas nuclei. 

\begin{figure}[ht]
	\centering
	\includegraphics[width=\columnwidth]{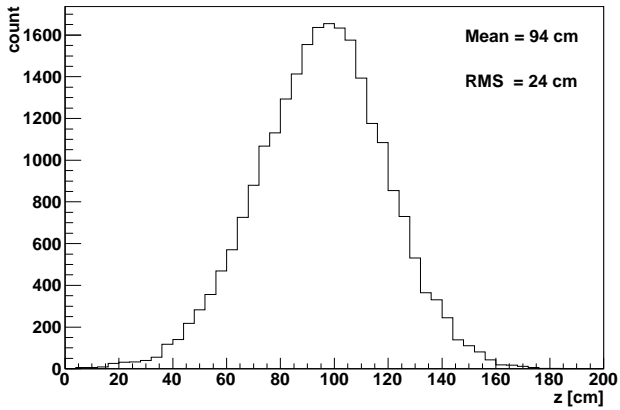}
	\caption{The distribution of maximum depth reached by the cooled muons in the cooling cell. \label{Zmax}}
\end{figure} 	
\begin{figure}[ht]
	\centering
	\includegraphics[width=\columnwidth]{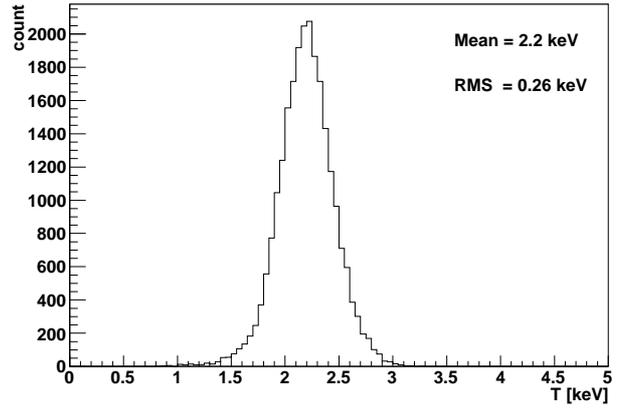}
	\caption{The energy distribution of the cooled muons directly before entering the window. \label{KE_Cell}}
\end{figure} 

	 The depth at which a muon turns around depends on its kinetic energy when it enters the  cooling cell. Figure\ \ref{Zmax} shows the distribution of this depth. The mean depth is \unit{94}{\centi\meter}. A \unit{2}{\meter} long cooling cell has sufficient length to turn back all the muons that can be cooled. The cooled muons have an equilibrium energy of \unit{2.2}{\kilo\electronvolt} with a spread of \unit{0.26}{\kilo\electronvolt} (Fig.\ \ref{KE_Cell}). 24\% of the muons from the source are cooled and reach the exit window. 
		
\subsection{Effect of the window}
	The window of the cooling cell is \unit{20}{\nano\meter} thick and made of carbon. 4\% of the cooled muons stop and decay in the window. The beam exits the window with a mean energy of \unit{0.9}{\kilo\electronvolt} at which $q_{\mathrm{eff}} = 0.15$. When the muons exit the window and enter the vacuum, a discrete charge is chosen based on $q_{\mathrm{eff}}$, which is interpreted here as the probability that the muon is not in muonium atom. 85\% of the cooled muons are therefore lost to muonium formation because the dipole does not turn the muonium atoms to the output channel.	
		
 \begin{figure}[ht]
	\centering
	\includegraphics[width=\columnwidth]{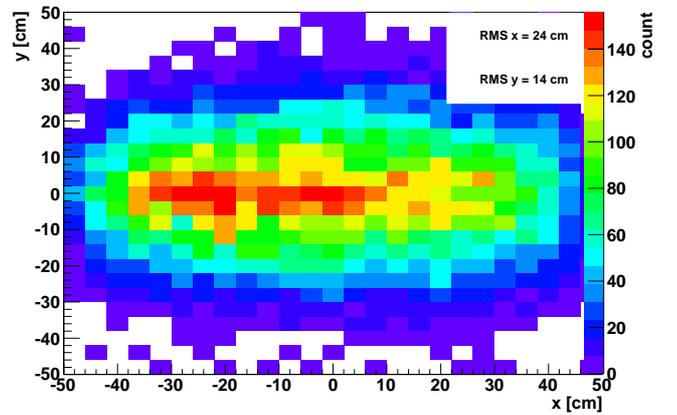}
	\caption{The spatial distribution of the cooled beam exiting the cooling cell (both \textmu\ensuremath{^{+}} and muonium). \label{XY_OutCell}}
\end{figure}
	The spatial distribution of the cooled muons exiting the cooling cell is shown in Fig.\ \ref{XY_OutCell}. The beam has a transverse RMS sizes of \unit{24}{\centi\meter} in $x$ and \unit{14}{\centi\meter} in $y$.
	
\subsection{Output} 	
\begin{figure}[ht]
	\centering
	\includegraphics[width=\columnwidth]{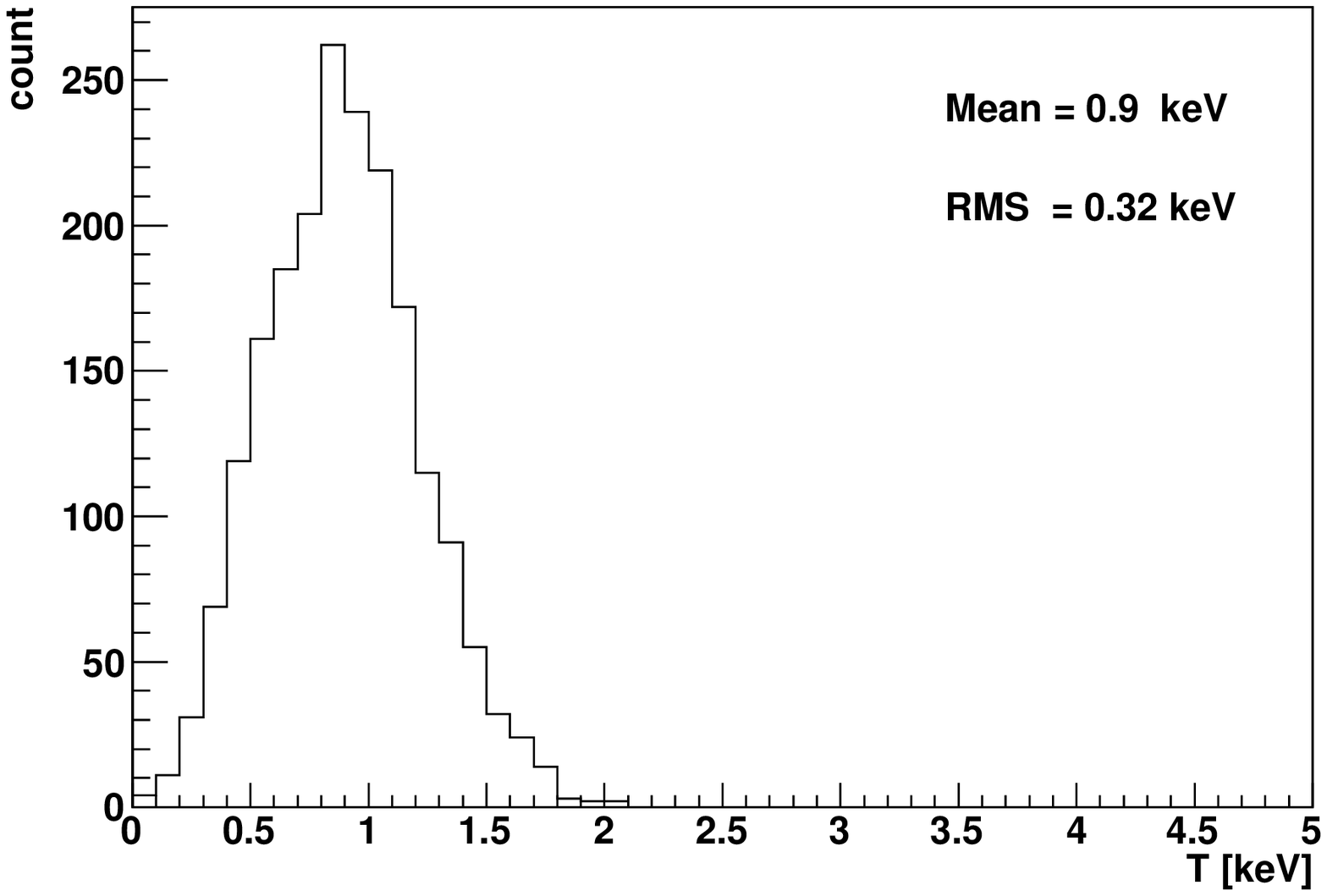}
	\caption{The energy distribution of the output \textmu\ensuremath{^{+}}. \label{KE_cool}}
\end{figure}
 \begin{figure}[ht]
	\centering
	\includegraphics[width=\columnwidth]{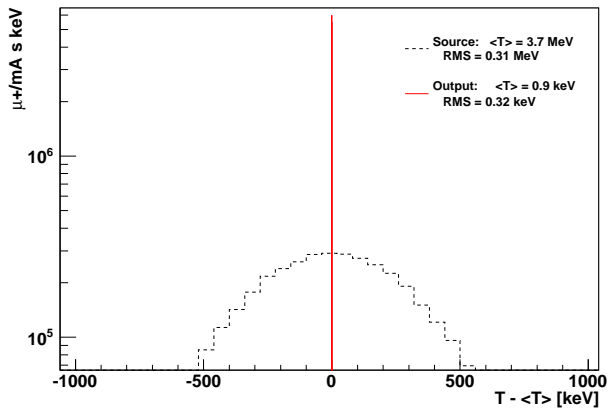}
	\caption{The energy distribution of the output beam compared to the input beam. The distributions are centered on their means. \label{output}}
\end{figure}
 \begin{figure}[ht]
	\centering
	\includegraphics[width=\columnwidth]{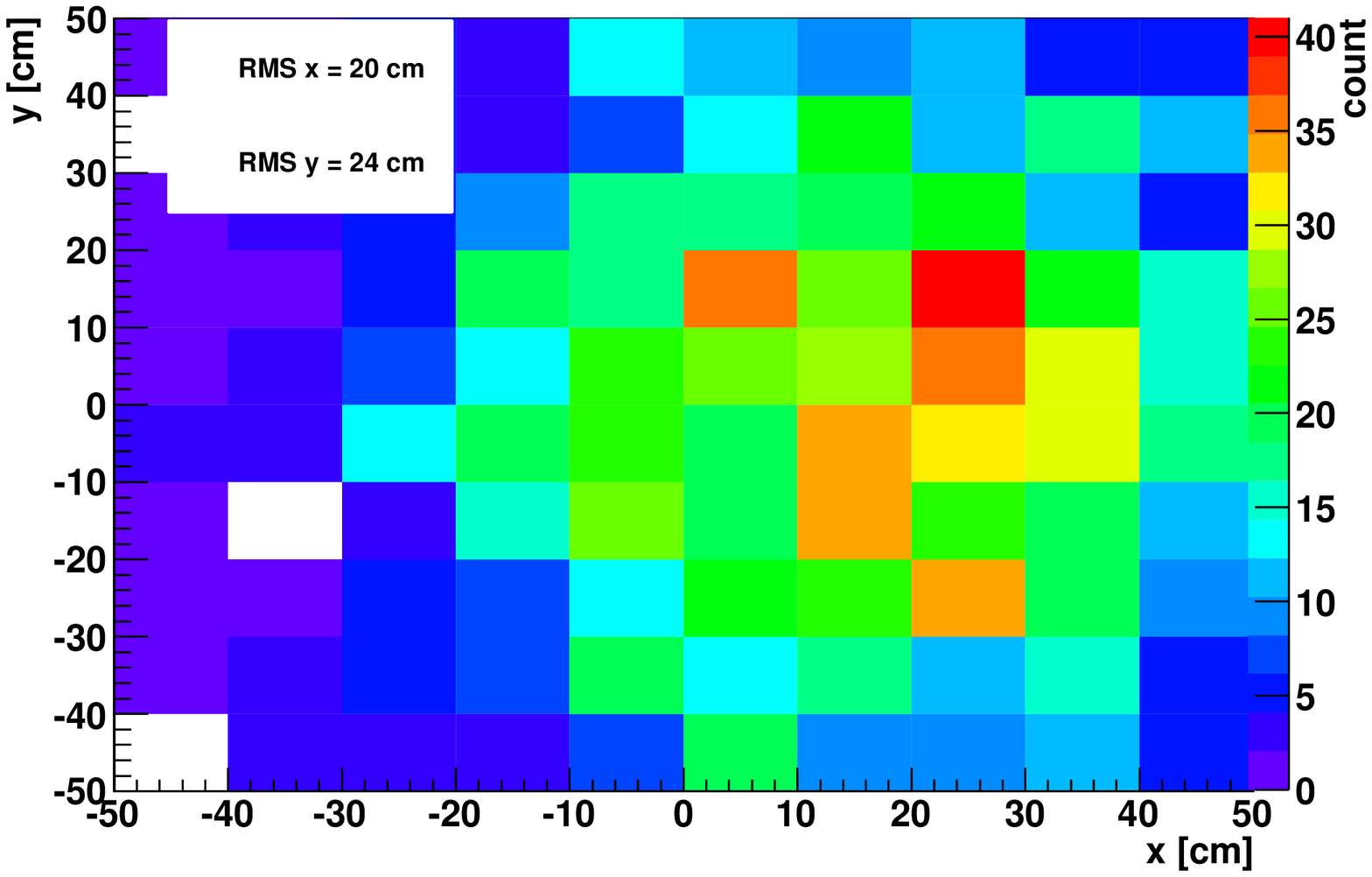}
	\caption{The transverse distribution of cooled muons exiting the dipole.\label{xy}}
\end{figure}
\begin{figure}[ht]
	\centering
	\includegraphics[width=\columnwidth]{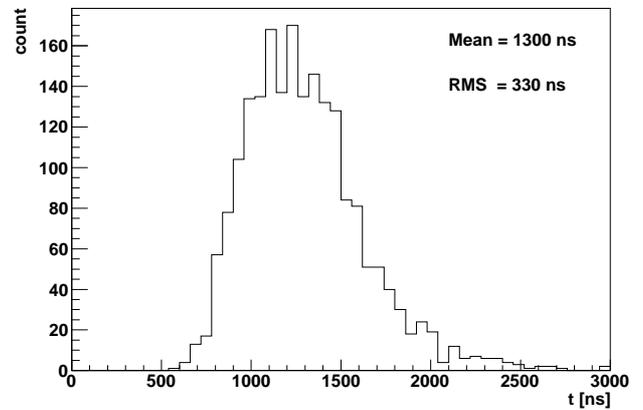}
	\caption{The distribution of time of flight of the cooled muons from source to the output.\label{time}}
\end{figure}

	After the dipole, the cooled beam has a mean kinetic energy of \unit{0.9}{\kilo\electronvolt} and an RMS of \unit{0.32}{\kilo\electronvolt} (Fig.\ \ref{KE_cool}). Figure~\ref{output} shows a comparison of the energy relative to the mean energy for muons at the source and muons exiting the scheme. The peak muon rate is \unit{6.0\times\power{10}{6}}{\per(\milli\ampere.\second.\kilo\electronvolt)} while the peak source rate is \unit{2.9\times\power{10}{5}}{\per(\milli\ampere.\second.\kilo\electronvolt)}. The energy spread of the beam is greatly reduced. The spatial distribution of muons exiting the dipole is shown in Fig.\ \ref{xy}. The RMS sizes of the beam are \unit{20}{\centi\meter} in $x$ and \unit{24}{ \centi\meter} in $y$. 2\% of the source muons exit the dipole; they would then be collected and  reaccelerated. From the  initial source to the output, the mean time of flight is \unit{1300}{\nano\second} (Fig.\ \ref{time}). 

\subsection{A simplified scheme}

 \begin{figure}[ht]
	\centering
	\includegraphics[width=\columnwidth]{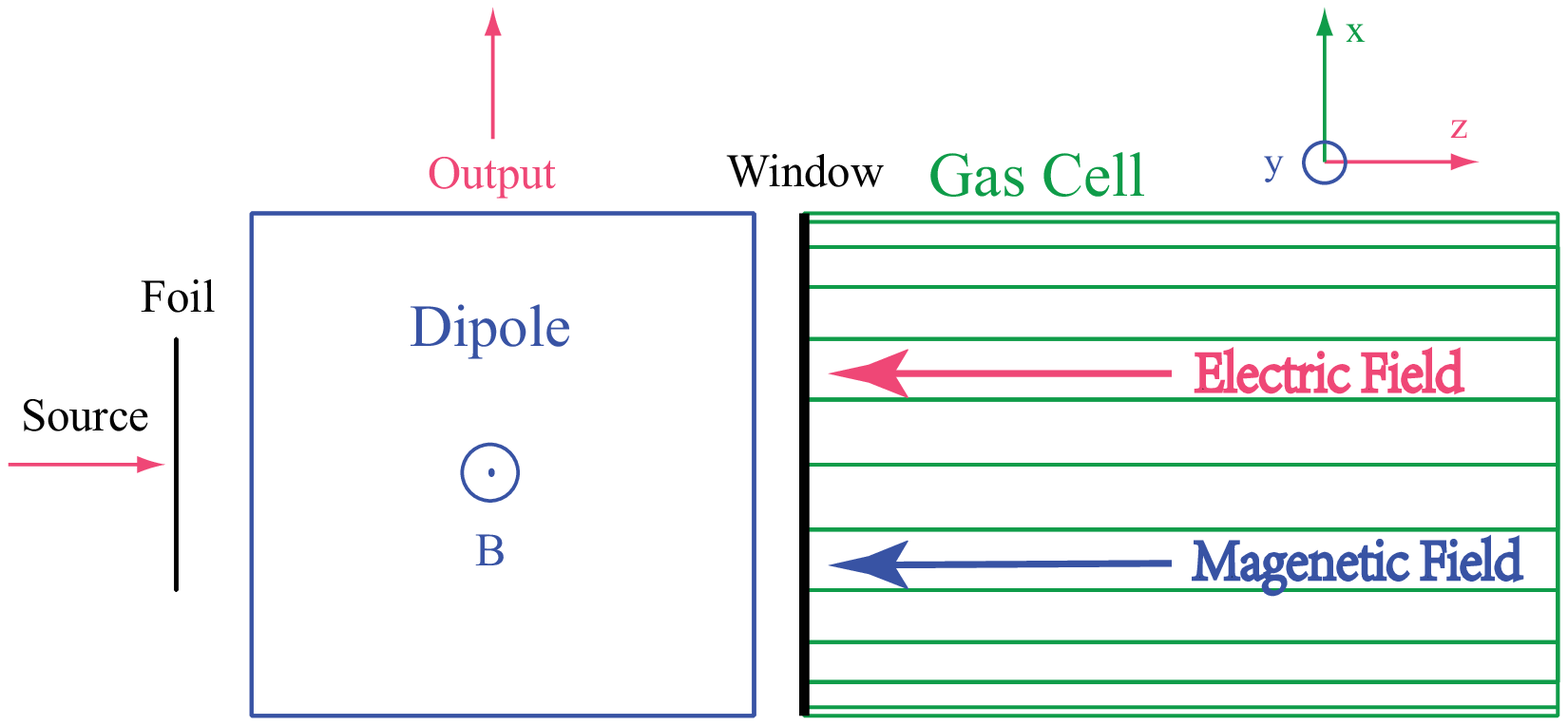}
	\caption{Block diagram of the simplified scheme.\label{simple_scheme}}
\end{figure}
	We also simulated a simplified scheme that avoids the complication of the quadrupoles and greatly reduces the required high voltage.

	We limited the design so that the required potential is within the available range of common high voltage supplies. The quadrupoles are removed but the dipole is kept for extracting the cooled muons. The diagram of this scheme is shown in Fig.\ \ref{simple_scheme}. A foil reduces the energy of the initial beam to maximize the number of muons that can be cooled by the cooling cell. The thickness of the foil for this scheme is \unit{130}{\micro\meter}. The dipole strength is \unit{0.01}{\tesla}, which turns the cooled muons with a radius of \unit{10}{\centi\meter}. The cooling cell is only \unit{30}{\centi\meter} long and has a diameter of \unit{20}{\centi\meter}. Keeping the electric field at \unit{1.8}{\mega\volt\per\meter} yields  a potential difference of \unit{540}{\kilo\volt} in the cooling cell. A \unit{1}{\tesla} solenoidal magnetic field guides the beam in the cooling cell. The cell is filled with helium gas at a density of \unit{0.01}{\milli\gram\per\centi\cubic\meter}. A \unit{20}{\nano\meter} thick carbon window separates the cooling cell from the dipole.
	
	At the output, the cooled beam has a mean kinetic energy of \unit{0.9}{\kilo\electronvolt} and an RMS of \unit{0.32}{\kilo\electronvolt}.  The overall efficiency in this scheme is 0.1\%. The peak muon rate is \unit{4.2\times\power{10}{5}}{\per(\milli\ampere.\second.\kilo\electronvolt)}.

\section{Points to investigate}
	This paper outlines a frictional cooling scheme for the efficient production of a LE-\textmu\ensuremath{^{+}} beam. There are several issues that require further study.
\subsection{Fields}
	All fields in the simulations are uniform. The fringe fields of a strong solenoid around the cooling cell would change the muons' trajectories. Bucking coils could be used to reduce the fringe fields out of the cooling cell. The dipole can be put further away from the cooling cell to avoid the fringe field of the solenoid; or, the fringe field itself can replace the dipole and guide the cooled muons out. Realistic scenarios need to be simulated.
	
	An additional radial electric field could focus the beam to the axis of the cell, where the radial component of the fringe magnetic field is much lower than at the edge of the solenoid.  How to create such a field  should be investigated.
	
\subsection{Electric fields in gas}
	The frictional cooling scheme requires a strong electric field in gas. Frictional cooling experiments at the Max-Planck-Institute for Physics have used electric fields up to \unit{500}{\kilo\volt\per\meter} over \unit{10}{\centi\meter} in a gas cell filled with helium at pressures from \unit{10^{-3}}{\milli\bbar} to \unit{1.25}{\bbar} without breakdown. Further investigations are needed to determine the feasibility of even higher fields in helium gas. 

	Muons lose a small fraction of their kinetic energy to ionization in the gas. We estimated an ionization rate of less than \unit{10^{10}}{\per\second} for a source flux of \unit{4.2\times 10^{8}}{\per\second}. The removal or neutralization of the resultant charge in the cooling cell will have to be investigated.
	
\subsection{Window}
	The window on the cooling cell must be thin and gas tight.  \unit{20}{\nano\meter} thin carbon films can be deposited on nickel grids with 55\% open area~\footnote{Structure Probe Inc.: http://www.2spi.com}. These windows can hold high gas pressures over small areas. However, they cut the efficiency of the scheme in half. In our case, this would mean 1\% overall efficiency. The feasibility of large-area thin gas-tight windows needs to be investigated. Focusing the beam in the cooling cell would be helpful by reducing the area required for the thin window. The material of the window should also be optimized to reduce beam neutralization. A layer of noble gas frozen onto the window can decrease neutralization of the cooled muons. Other technologies such as differential pumping systems could be considered to replace the window.
	
\section{Conclusion}
	Frictional cooling holds promise for producing low energy muon beams with small energy spreads. Our simulation shows that a frictional cooling scheme can cool a surface muon beam to a mean energy of \unit{0.9}{\kilo\electronvolt} with an energy spread of \unit{320}{\electronvolt}. The efficiency for LE-\textmu\ensuremath{^{+}} production is greater than 1\%, which is 2 to 3 orders of magnitude higher than the efficiency achieved with the moderation technique. If the frictional cooling scheme is experimentally successful, a surface muon rate of \unit{228\times\power{10}{6}}{\per(\milli\ampere.\second)} would result in a LE-\textmu\ensuremath{^{+}} rate of \unit{2.3\times\power{10}{6}}{\per(\milli\ampere.\second)}.	

\bibliographystyle{elsart-num}
\bibliography{LEmuViaFC}
	

\end{document}